\documentclass{emulateapj} 
\usepackage{apjfonts}

\newcommand\lsim{\lesssim}
\newcommand\gsim{\gtrsim}

\begin{document}

\title{UV pumping of hyperfine  transitions in the light elements, with application to 21-cm hydrogen and 92-cm deuterium lines from the early universe}
\author{Leonid Chuzhoy and Paul R. Shapiro}
\bigskip
\affil{McDonald Observatory and Department of Astronomy, The University of Texas at Austin, RLM 16.206, Austin, TX 78712, USA; chuzhoy@astro.as.utexas.edu; shapiro@astro.as.utexas.edu}

\begin{abstract}
We present new analytic calculations of the coupling between
ultraviolet resonance photons and the population of the hyperfine
states in the light elements (H, D, ${\rm ^3He^+}$) which include several
previously neglected physical processes. Among these are the backreaction of resonant scattering on the pumping radiation, the scattering of Ly$\beta$ photons and the effect of local departure from pure Hubble flow. The application of the new
treatment to the redshifted hydrogen 21 and deuterium 92 cm lines
from the high-redshift universe results in an amplitude correction
of up to an order of magnitude. We further show that the standard
assumption that ultraviolet pumping drives the spin temperature
towards the kinetic temperature does not hold for deuterium, whose
spin temperature is generally negative.
\end{abstract}

\keywords{cosmic microwave background -- cosmology: theory -- diffuse radiation -- intergalactic medium -- radio lines: general}

\section{\label{Int}Introduction}
The hyperfine transitions in hydrogen, deuterium and helium-3 provide an opportunity to study the universe during its ``dark ages'' and the epoch of reionization that followed. From the radio continuum produced by redshifted lines of these elements, which can be observed in absorption or in emission against the cosmic microwave background (CMB),  information about the thermal evolution, density perturbation spectrum and ionization history of the universe can be obtained. However, the correct interpretation of the measured signal requires knowledge of the processes which decouple the spin temperature of the atoms, $T_{\rm s}$, from the CMB temperature, $T_{\rm CMB}$. Of these, one of the most important during the epoch of reionization ($6\lsim z\lsim 20$) is ultraviolet pumping. Resonant photons, such as Ly$\alpha$, absorbed by the atom in the ground state may, after the decay of the excited state, leave the electron in a different hyperfine level. This process couples the spin temperature to the color temperature of the resonant photons, $T_\alpha$ \cite{f8}
\begin{eqnarray}
\label{Tsp}
T_{\rm s}=\frac{T_{\rm CMB}+y_\alpha T_\alpha+y_{\rm c}T_{\rm k}}{1+y_\alpha+y_{\rm c}},
\end{eqnarray}
where $T_{\rm k}$ is the kinetic temperature, $y_{\rm c}$ is a constant proportional to the collisional excitation rate and $y_\alpha$ is given by
\begin{eqnarray}
y_{\alpha}=\frac{P_{10}T_*}{A_{10}T_{\alpha}}
\end{eqnarray}
where $T_*=h\nu_{\rm hyp}/k$, $\nu_{\rm hyp}$ is the frequency of the hyperfine transition, $A_{10}$ is the spontaneous emission coefficient of the hyperfine transition and $P_{10}$ is the radiative deexcitation rate for the upper hyperfine level due to Ly$\alpha$ photons. 

As we show in this paper, the correct application of the above expressions requires us to take into account several physical processes which have previously been neglected. 
The first of these is the effect of the hyperfine splitting on the photon color temperature and intensity. In their pioneering papers, Wouthuysen (1952) and Field (1958, 1959) argued that in the limit of large optical depth,  $T_\alpha$  must approach the kinetic temperature of the atoms, to which the photons are coupled by recoil. 
As a result, since the  Ly$\alpha$ resonance scattering depth for hydrogen atoms in the neutral intergalactic medium (IGM) is extremely large, it has been customary ever since then to replace $T_\alpha$ with $T_{\rm k}$ in eqs. 1 and 2 above.
That argument, however, did not take into account the effect of the hyperfine splitting on the change of photon energy during scattering.  Just as the spin temperature of the atoms is affected by the spectrum of the UV photons in the resonance line, so the UV spectrum around the resonance is in turn affected by the spin temperature. This backreaction shifts $T_\alpha$ towards $T_{\rm CMB}$, thus reducing the coupling between the kinetic and spin temperatures. In addition, we consider the effect of scatterings on the radiation intensity near the resonance. Chen \& Miralda-Escude (2004) showed that, in a homogeneously expanding cosmological medium, the intensity of the continuum radiation drops around the resonance, thus reducing the value of $y_\alpha$. Here we derive an analytical formula for the amplitude of this drop. We find significant discrepancies between our analytical calculations and some of their numerical results. Furthermore, we extend their analysis to consider departures from pure Hubble flow, associated with cosmic structure formation, such as a contracting medium which is appropriate for overdense regions. Finally, we also consider scattering by resonant photons other than Ly$\alpha$. Ly$\alpha$ photons have the highest scattering cross-section and, unlike higher resonances like Ly$\beta$, they are not destroyed by cascade after just a few scatterings. For these reasons, the higher resonances are generally assumed to be unimportant. While correct for hydrogen, this assumption proves to be a very gross error in the case of deuterium, for which the role of Ly$\beta$ photons is crucial. 

The corrections derived here to the hydrogen spin temperature which results from Ly$\alpha$ pumping in the early universe can have a significant effect on predictions of the 21-cm brightness temperature and its fluctuations.  As we shall show their importance depends on the intensity of the UV background and the temperature of the IGM. An approximation which is often made in recent theorectical predictions of the 21-cm background from the epoch reionization involves the assumption that the UV pumping background is high enough to make $T_s$ equal $T_k$, while some other process (e.g. X-ray background) heats the IGM without ionizing it, to $T_k\gg T_{CMB}$ \cite{MMR,T,C,CM,ZFH,FSH,BL,Aal,Sal,Mal}. In that limit the 21-cm brightness temperature becomes independent of the actual value of $T_s$, so the calculations are greatly simplified. However, the period either before the UV background reaches such high intensity or the neutral IGM is heated is not well represented by this limiting approximation. This includes the epoch when the 21-cm signal is in absorption and potentially at its strongest. That is the case in which the results presented here will be important. 

While this paper was in preparation, a preprint of Hirata (2006) was posted in which some of the above effects were calculated numerically.
We find an excellent agreement between our analytic solutions and his numerical results. 

 In \S 2, we estimate the effect of the hyperfine transition on the color temperature of Ly$\alpha$ photons.
In \S 3, we calculate the change of intensity of Ly$\alpha$ photons due to their scattering by atoms.
In \S 4, we estimate the effect of these processes on the radio emission signals from D and ${\rm ^3He^+}$. In \S 5, we discuss the implications of our results for the interpretation of hydrogen and deuterium radio signals from the epoch of reionization and the end of the cosmic dark ages.

\section{The coupling between the UV and radio spectra}
Consider a UV resonance transition which is followed by a single-photon decay to either hyperfine level of the ground state.
Upon such scattering by an atom, the energy of the incoming UV photon is changed due either to atomic recoil or to the hyperfine transition. In most cases, both the kinetic and the spin temperature of the atoms is below the energy of the photons ($h\nu/k\sim10^5$ K), so on average the photons will lose energy in scatterings. The actual energy change, however, depends on the exact frequency of the photon. Assuming that the energy distribution of atoms is Maxwellian, we found that, in each scattering, the average energy loss of the photons is (see Appendix for details)
\begin{eqnarray}
\Delta E_1=\frac{(h \nu)^2}{mc^2}\left(1-\frac{kT_{\rm k}}{h}\frac{\sigma'(\nu)}{\sigma(\nu)}\right),
\end{eqnarray}
due to the atomic recoil and
\begin{eqnarray}
\Delta E_2= \frac{b(h \nu_{\rm hyp})^2}{2kT_{\rm s}}\left(1-\frac{kT_{\rm s}}{h}\frac{\sigma'(\nu)}{\sigma(\nu)}\right),
\end{eqnarray}
due to the change of the hyperfine state, where $\sigma(\nu)$ is the scattering cross-section, $\sigma '(\nu)=d\sigma/d\nu$, $\nu_{\rm hyp}$ is the frequency of the hyperfine transition, $b$ is the probability that a single resonance scattering results in a spin-flip and $m$ is the atomic mass. Similarly, the total energy loss in the two processes may be written as
\begin{eqnarray}
\Delta E=\Delta E_1+\Delta E_2=\left(A-B\frac{k}{h}\frac{\sigma'(\nu)}{\sigma(\nu)}\right),
\end{eqnarray}
where $A=(h^2 \nu^2/mc^2)+(bh^2\nu_{\rm hyp}^2/2kT_{\rm s})$ and $B=(T_{\rm k}h^2 \nu^2/mc^2)+(bh^2\nu_{\rm hyp}^2/2k)$.

Consider the effect on the incoming radiation spectrum of repeated scatterings like those described above. This is relevant whenever the incoming photon is scattered without the possibility of photon destruction by cascade.
For the radiation spectrum, increasing $A$ and $B$ by the same factor produces the same effect as an increase in the optical depth. Thus the color temperature of the resonant photons, which approaches a constant in the limit of high optical depth, can depend only on the ratio of $A$ and $B$.  Considering either the recoil or the hyperfine transition alone would give us either $B/A=T_{\rm k}$ or $B/A=T_{\rm s}$, respectively, so obviously $B/A=T_\alpha$. For hydrogen Ly$\alpha$, $b=2/9$, and  the color temperature is given by
\begin{eqnarray}
\label{Tcol} T_{\alpha}=\frac{(1+2.5T_{\rm k})T_{\rm s}}{1+2.5T_{\rm
s}},
\end{eqnarray}
where all temperatures are in units of degrees Kelvin. Predictably the value of $T_{\alpha}$ is between $T_{\rm k}$ and $T_{\rm CMB}$. Solving equations (\ref{Tsp}) and (\ref{Tcol}) then gives the correct spin temperature
\begin{eqnarray}
\label{Tf}
T_{\rm s}=\frac{T_{\rm CMB}+[y_{\alpha,eff}+y_{\rm c}]T_{\rm k}}{1+y_{\alpha,eff}+y_{\rm c}},\\
\label{yeff}
y_{\alpha,eff}=\left[\frac{P_{10}T_*}{A_{10}T_{\rm k}}\right]\left[1+0.4T_{\rm k}^{-1}\right]^{-1}.
\end{eqnarray}
The the result in eq. \ref{Tf} is identical to the commonly used approximation for $T_{\rm s}$ described in \S 1, except that the value of the coupling constant $y_{\alpha}$ is reduced by a factor $(1+0.4T_{\rm k}^{-1})$. In the next section we will show that the radiative deexcitation rate, $P_{10}$, is also affected by the changing spectrum.

\section{The intensity of  Ly$\alpha$ photons}
\subsection{Expanding medium}
The  Ly$\alpha$ pumping photons can be divided into two groups according to their origin. The first group  (``continuum photons'') consists of continuum photons originally emitted in the frequency range between the Ly$\alpha$ and Ly$\beta$ transitions, which, due to the cosmological redshift, are eventually shifted into the Ly$\alpha$ resonance.
The second group (``injected photons'') owes its origin to photons originally emitted between Ly$\gamma$ and Ly-limit frequencies. When these photons redshift into one of the Lyman series resonances, they get absorbed by atoms, which are thereby excited to $n>3$ state. 
Since the subsequent cascade mostly likely goes directly to the ground state, most photons are reemitted close to their original frequency (with a small change due to a Doppler shift) and are soon reabsorbed. However, after just a few scatterings most of the original photons are destroyed by splitting into several photons. Roughly two thirds of all cascades from $n>3$ states and all off the cascades from $n=3$ state, go through the intermediate $2s$ state, which decays by two-photon emission. The rest pass through the $2p$ state, whose decay produces a Ly$\alpha$ photon. Unlike the ``continuum photons'', which reach the Ly$\alpha$ resonance from the blue wing, the``injected photons'' are produced directly in its center.

When the Ly$\alpha$ optical depth is high, the spectrum of the photons near the Ly$\alpha$ resonance can be found using the Fokker-Planck approximation. In the comoving frame of the gas, the intensity $J(\nu)$ varies with frequency in the neighborhood of the resonance at $\nu_\alpha$ according to the following (see Appendix B for derivation)
\begin{eqnarray}
\label{J}
 \phi(x)J'(x)+2[\eta \phi(x)+\gamma ]J(x)=2\gamma J_0(1-k_\alpha \Theta(x)),
\end{eqnarray}
where $x=(\nu/\nu_\alpha-1)/(2kT_{\rm k}/mc^2)^{1/2}$ is the frequency distance from the line center in dimensionless units, $\phi(x)$ is the normalized absorption cross-section, $J_0$ is the intensity on the far red side of the resonance (i.e., where it is unaffected by scattering) of either the continuum or injected photons, and $k_\alpha$ equals zero for continuum or unity for injected photons. 

\begin{figure}
\resizebox{\columnwidth}{!}
{\includegraphics{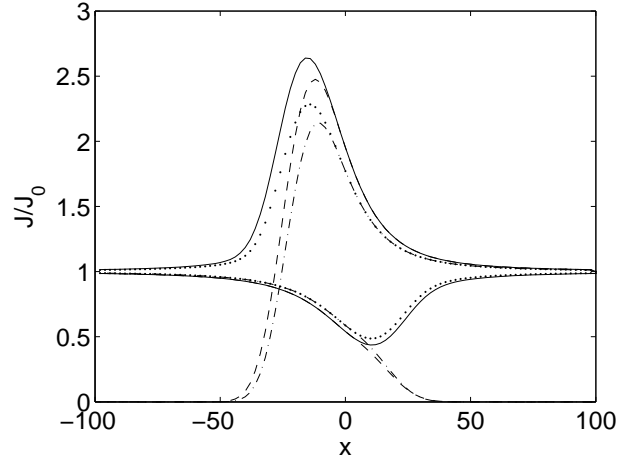}}
\caption{\label{JH}
The spectrum profile around hydrogen Ly$\alpha$ resonance in an expanding region with $\tau_{\rm GP}=10^{6}$ and $T_{\rm k}=3$ K and in a contracting region with  $\tau_{\rm GP}=3\cdot 10^{7}$ and $T_{\rm k}=10$ K. The solid and dashed curves are for $T_{\rm s}=T_{\rm k}$, for continuum and injected photons respectively. The dotted and dashed-dotted lines are for $T_{\rm s}=T_{\rm CMB}=35$ K, for continuum and injected photons respectively.}
\end{figure}

If hyperfine transitions are neglected, the recoil parameter, $\eta$, equals $h\nu_\alpha/(2kT_{\rm k}mc^2)^{1/2}$ \cite{B} and $\gamma^{-1}=\tau_{\rm GP}$ is the Gunn-Peterson optical depth to Ly$\alpha$ resonance scattering. Including the effect of the hyperfine transition gives a correction of order $(1+0.4/T_{\rm k})$ to the values of  $\gamma$ and  $\eta$ (see Appendix B).  Generally, however, this correction results in just a few per cent change in the radiation intensity (see Fig. \ref{JH}).
 Neglecting terms of order $\gamma$ , the solution of equation (\ref{J}) can be written as \footnote{It has been recently pointed to us that a similar solution for an expanding medium was found by Grachev(1989).}
\begin{eqnarray}
\label{J1}
J(x)=J(0)e^{-\frac{2 \pi  \gamma  x^3}{3 a}-2 \eta  x},  \hspace{0cm}  
{\rm \; for \; k_\alpha=1\; and \; x>0}, \hspace{1cm}  \nonumber \\
=J_0  e^{-\frac{2 \pi  \gamma  x^3}{3 a}-2 \eta  x} \int_{-\infty }^x 2\pi  \gamma a^{-1}e^{\frac{2 \pi  \gamma 
   z^3}{3 a}+2 \eta  z}  z^2 \,
   dz,  
    \; {\rm otherwise}, \hspace{0cm} 
\end{eqnarray}
where $a=A_{21}(2kT_{\rm k}/mc^2)^{-1/2}/4\pi\nu_\alpha$ and $A_{21}$ is the Einstein spontaneous emission coefficient of the Ly$\alpha$ transition. 
When the optical depth is low or the temperature is high, so that thermal width of the line becomes similar to the width of absorption feature, 
the approximations we used in deriving equations (\ref{J})-(\ref{J1}) break down and higher order terms become more important \cite{R6,FP}. However, under such conditions the effect of  Ly$\alpha$ scattering on the radiation intensity is in any case negligible and the equation (\ref{J1}) still yields the correct answer. 

The kinetic temperature of the gas, $T_{\rm k}$, is coupled to  $T_{\rm CMB}$ by inverse Compton scatterings at redshifts $z\gsim 150$. Thereafter, until the IGM is reheated during the epoch of reionization, the gas temperature falls adiabatically, according to $T_{\rm k}\approx 0.02(1+z)^2$. Prior to reionization, the optical depth of the neutral hydrogen evolves according to 
\begin{eqnarray}
\tau_{\rm GP}=7\cdot 10^5 \left(\frac{\Omega_{b} h_{100}}{0.03}\right) \left(\frac{\Omega_{m}}{0.25}\right)^{-1/2} \times \nonumber \\
\left(\frac{1+z}{10}\right)^{3/2}|\frac{H'}{H}|^{-1}(1+\delta),
\end{eqnarray}
 where $H'$ and $\delta$ are the local values of the Hubble constant (to account for departure from unperturbed Hubble expansion) and the overdensity, respectively.
Figure \ref{JH} shows an example of the spectrum profile, $J(x)$, for conditions corresponding to unheated and unperturbed neutral hydrogen at $z\sim 12$. Predictably, near the center of the line, where frequent scatterings with atoms make photons lose energy faster, $J(x)$ shows a dip, whose amplitude is the same for the continuum and injected photons
\begin{eqnarray}
\label{J3}
\frac{J(0)}{J_0}=\frac{\pi  \zeta  \left(J_{\frac{1}{3}}(\zeta
   )-J_{-\frac{1}{3}}(\zeta )\right)}{\sqrt{3}}+ 
   _1F_2\left(1;\frac{1}{3},\frac{2}{3};-\frac{\zeta
   ^2}{4}\right),
\end{eqnarray}
where $\zeta=(16\eta^3a/9\pi\gamma)^{1/2}$ (for hydrogen $\zeta=4.6\cdot 10^{-4}\gamma^{-1/2}T_{\rm k}^{-1}$), $_1F_2$ is a hypergeometric function and $J_{\frac{1}{3}}$ and  $J_{-\frac{1}{3}}$ are the Bessel functions of the first kind. Conveniently $J(0)/J_0$ has the same value for continuum and injected photons.

\begin{figure}
\resizebox{\columnwidth}{!}
{\includegraphics{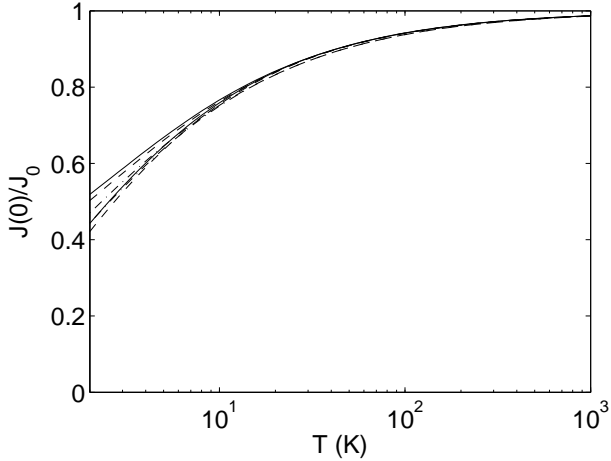}}
\caption{\label{figH} $J(0)/J_0$ at $z=12$ for $\delta=0$. The solid, dashed and dashed-dotted lines are obtained using respectively, eqs.  \ref{J3} and \ref{J4}, and the numerical fit of Hirata (2005) for $T_s=T_{CMB}$ (upper lines) or $T_s=T_k$ (lower lines, where solid and dashed-dotted lines in this case are indistinguishable).}
\end{figure}

 Typically, $\zeta$ is well below unity, which allows us to approximate the eq. \ref{J3} by
\begin{eqnarray}
\label{J4}
\frac{J(0)}{J_0}=e^{-1.69\zeta^{2/3}}.
\end{eqnarray}
Since most of the scatterings happen around $x=0$, the overall scattering rate and $y_\alpha$ fall by the same factor, $e^{-1.69\zeta^{2/3}}$. 
Hirata (2006) found this factor numerically, separately using a Fokker-Planck approximation and Monte Carlo simulations.  
Figure \ref{figH} shows excellent agreement between our equations \ref{J3} and \ref{J4}, and Hirata's numerical fit. 

\begin{figure}
\resizebox{\columnwidth}{!}
{\includegraphics{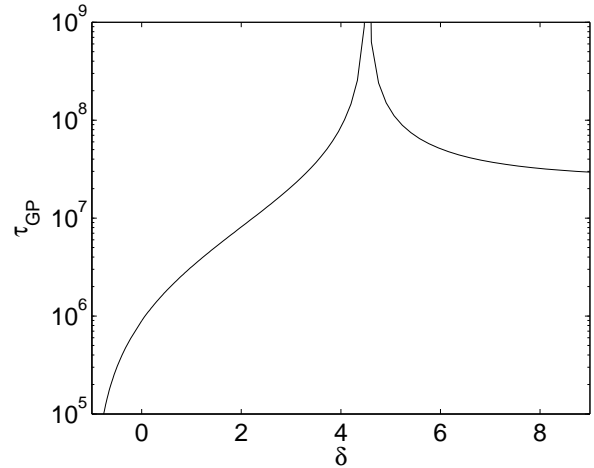}}
\caption{\label{Ga}Hydrogen optical depth $\tau_{\rm GP}$ for Ly$\alpha$ scattering at $z=12$ vs overdensity, $\delta=(\rho/<\rho>)-1$. \vspace{0.cm}}
\end{figure}

\subsection{Contracting medium}
Equation (\ref{J}) may after a slight change be used for a contracting, rather than expanding, medium, according to
\begin{eqnarray}
\label{J11}
\phi(x)J'(x)+2[\eta \phi(x)-\gamma]J(x)=2\gamma J_0(k_\alpha\Theta(-x)-1). \hspace{0.5cm}
\end{eqnarray}
In a contracting gas, the energy loss from scatterings slows down the blueshifting of the photons. Therefore instead of falling $J(x)$ actually rises {\it above} $J_0$ at the center, according to
\begin{eqnarray}
\label{J2}
\frac{J(0)}{J_0}=2 a^2 \eta ^2\, _1F_2\left(1;\frac{4}{3},\frac{5}{3};\frac{\zeta
   ^2}{4}\right) + \hspace{2cm} \nonumber \\
\sqrt[3]{\frac{2}{3}} \pi  \zeta
   ^{2/3} Bi\left(\left(\frac{3 \zeta }{2}\right)^{2/3}
  \right)+\,
   _1F_2\left(1;\frac{1}{3},\frac{2}{3};\frac{\zeta
   ^2}{4}\right),
\end{eqnarray}
where $Bi$ is the Airy function. 
Once again, this result applies to both continuum and injected photons.

\section{The rare isotopes}
\subsection{Deuterium}
When absorption of a Ly$\beta$ photon excites a hydrogenic atom to the 3p state, the decay proceeds either directly to the 1s state, in which case another Ly$\beta$ photon is emitted, or via 2s state, in which case the original Ly$\beta$ photon is split into three photons.
At redshifts $z\lsim 20$, the optical depth for deuterium Ly$\beta$ photons is below $10$. Since only 1 in $\sim 8$ of the absorbed photons is destroyed by three-photon reemission, the ratio of Ly$\beta$ to  Ly$\alpha$ scatterings does not drop as drastically for deuterium as for hydrogen, whose Ly$\beta$ optical depth is much higher. Still, this ratio is rather low ($\sim 0.1$ if the UV continuum spectrum is nearly flat between Ly$\alpha$ and Ly$\gamma$), so Ly$\beta$ photons would not have been important for deuterium either, if their color temperature were determined by the same process as for Ly$\alpha$ . Unlike Ly$\alpha$ photons, however, whose energy is changed only slightly by the scatterings and whose spectrum near the resonance is relatively flat, the spectrum around the Ly$\beta$ resonance is determined predominantly by photon destruction. As photons are redshifted across the resonance, more and more of them are destroyed, thus making the red wing of the resonance much weaker than the blue.

Since the Ly$\beta$ optical depth of deuterium is not very high, the Fokker-Planck approximation can no longer be used, and the integral equation must be solved instead.
For an isotropically expanding medium, the transfer equation for radiation in a steady state takes the form 
\begin{eqnarray}
\phi(x)J(x)-\gamma J'(x)=\int R(x,x')J(x')dx',
\end{eqnarray}
where $R(x,x')$ is the redistribution function \cite{RD}. For the photons which survive the scattering (i.e., for those which are not destroyed by cascade) $R(x,x')={\rm Erfc}(Max[|x'|,|x|])/2$ \cite{Un}. Taking into account the fraction of the photons that are destroyed, $f_{\rm d}$, gives the full redistribution function $R(x,x')=(1-f_{\rm d}){\rm Erfc}(Max[|x'|,|x|])/2$. For moderate optical depth the normalized absorption cross-section can be approximated as $\phi(x)=e^{-x^2}/\sqrt{\pi}$. Using this we obtain
\begin{eqnarray}
\label{D}
\frac{e^{-x^2}}{\sqrt{\pi}}J(x)-\gamma J'(x)= \hspace{3cm} \nonumber \\
\frac{1}{2}\int_{-\infty}^{\infty}(1-f_{\rm d}){\rm Erfc}(Max[|x'|,|x|])J(x')dx'.
\end{eqnarray}

\begin{figure}
\resizebox{\columnwidth}{!}
{\includegraphics{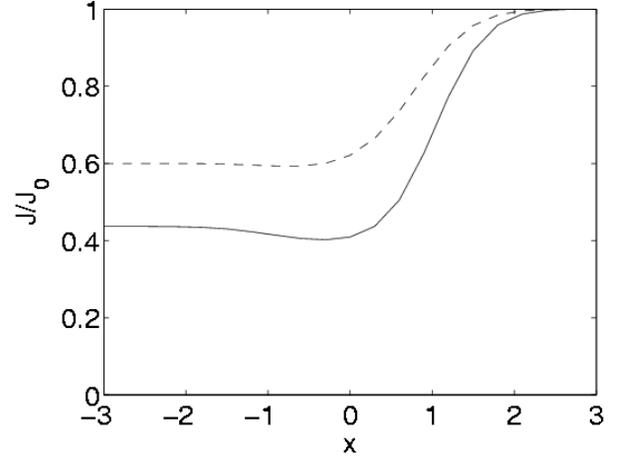}}
\caption{\label{JD}The spectrum profile around deuterium Ly$\beta$ resonance for $\gamma=0.1$ (solid line) and $0.2$ (dashed line).}
\end{figure}

Solving equation (\ref{D}) numerically (see fig. \ref{JD}),  we found that, around the Ly$\beta$ resonance, the color temperature is $T_\beta\sim (-0.2 T_{\rm k}^{1/2})$ K.

If the UV radiation field is sufficiently strong that we can ignore both the CMB photons and collisional coupling, the spin temperature is determined by the relative numbers of scatterings of Ly$\alpha$ and Ly$\beta$ photons with deuterium atoms (the combined effect of Ly$\gamma$ and higher resonances is generally a few times smaller than that of Ly$\beta$).
Taking into account the higher scattering cross-section of Ly$\alpha$ photons and the decrease of Ly$\beta$ intensity at the resonance, we obtain
\begin{eqnarray}
T_{\rm s}\sim -1 {\rm K}\; T_{\rm k}^{1/2}n_{\alpha\beta},
\end{eqnarray}
where $n_{\alpha\beta}$ is the ratio of Ly$\alpha$ to Ly$\beta$ photons. For hot radiation sources, which produce most of the UV photons, $n_{\alpha\beta}$ is close to the ratio of the bandwidths producing the photons $(\nu_{Ly\beta}-\nu_{Ly\alpha})/(\nu_{Ly\gamma}-\nu_{Ly\beta})\sim 3$. Thus deuterium can appear in emission with respect to the CMB radiation even if its kinetic temperature is below the CMB temperature.

\subsection{${\rm ^3He^+}$}
The case of ${\rm ^3He^+}$ emission is a relatively simple one. Since the corrections to radiation intensity and color temperature that were estimated in \S 2 and \S 3 are important only at low kinetic temperatures, we can neglect them for ionized helium which becomes abundant only at $T>10^3$ K. Furthermore, unlike deuterium,  ${\rm ^3He^+}$ resonances are to the red of ${\rm ^4He^+}$. Thus practically all resonant photons except Ly$\alpha$ are prevented from reaching ${\rm ^3He^+}$, being instead absorbed and destroyed by  ${\rm ^4He^+}$ ions first.

\section{Discussion}

We have shown that the accurate modeling of the photon spectrum around the hydrogen Ly$\alpha$ resonance results in a significant correction to the radiative pumping efficiency. For the cosmological parameters given by the latest WMAP results, $\Omega_{b0} h_0=0.03$ and $\Omega_{m0}=0.25$, \cite{Sp} and unperturbed Hubble expansion, $T_{\rm s}$ is correctly given by eq. \ref{Tf} if $y_{\alpha,eff}$ is given by
\begin{eqnarray}
\label{cor}
\frac{y_\alpha,eff}{y_{\alpha,0}}=e^{-0.37(1+z)^{1/2}T_{\rm k}^{-2/3}}\left(1+\frac{0.4}{T_{\rm k}}\right)^{-1},
\end{eqnarray}
where $y_{\alpha,0}$ is the previously used coupling constant, given by
\begin{eqnarray}
y_{\alpha,0}=\frac{16\pi^2 T_*e^2 f_{12}J_0}{27 A_{10}T_{\rm k} m_{\rm e}c},
\end{eqnarray}
where $f_{12}=0.416$ is the oscillator strength of the Ly$\alpha$ transition and $J_0$ is the intensity at Ly$\alpha$ resonance, when the backreaction caused Ly$\alpha$ scattering by is neglected . 

Prior to reionization the mean differential brightness temperature of the hydrogen 21 cm signal is given by
\begin{eqnarray}
\delta T_{\rm b}=0.03\; {\rm K} \left(\frac{T_{\rm s}-T_{\rm CMB}}{T_{\rm s}}\right) \times \hspace{2cm} \nonumber \\
\left(\frac{\Omega_{b} h_{100}}{0.03}\right) \left(\frac{\Omega_{m}}{0.25}\right)^{-1/2}  \left(\frac{1+z}{10}\right)^{1/2}.
\end{eqnarray}
From the above equation it follows that the strongest signal is
obtained when the spin temperature is well below $T_{\rm CMB}$, in
which case $T_{\rm b}\propto T_{\rm s}^{-1}$. When radiation sources
begin to form, $T_{\rm s}$ indeed starts to drop as it decouples
from $T_{\rm CMB}$ and approaches $T_{\rm k}$ instead. However, if
X-ray photons are present, with an intensity which increases simultaneously with those in the UV, these
would at the same time heat the gas, so that, eventually, $T_{\rm k}$
and $T_{\rm s}$ become greater than $T_{\rm CMB}$.  Chen \&
Miralda-Escude (2004) estimated that in this case the strongest signal ($\delta T_{\rm
b}\sim -20$ mK) arises when the gas is still in absorption and $T_{\rm s}$ is not yet fully
decoupled from $T_{\rm CMB}$, i.e., when the signal is still very
sensitive to the changes in the coupling strength,  $y_{\alpha}$. 

\begin{figure}
\resizebox{\columnwidth}{!}
{\includegraphics{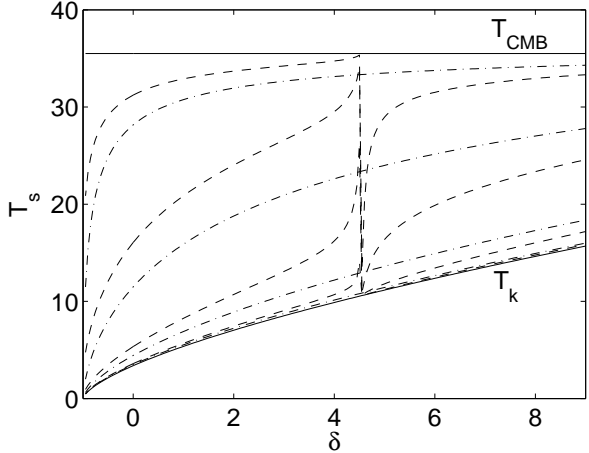}}
\caption{\label{T1}Hydrogen spin temperature at $z=12$ for different radiation intensities, $y_{\alpha,0} T_{\rm k}=1$, 10, $10^2$, $10^3$ K (lower curves correspond to higher values of  $y_{\alpha,0}$). 
The dashed and the dashed-dotted lines are calculated, respectively, with and without the correction to the coupling constant.
The solid lines are $T_{\rm CMB}$ and $T_{\rm k}$.}
\end{figure}

\begin{figure}
\resizebox{\columnwidth}{!}
{\includegraphics{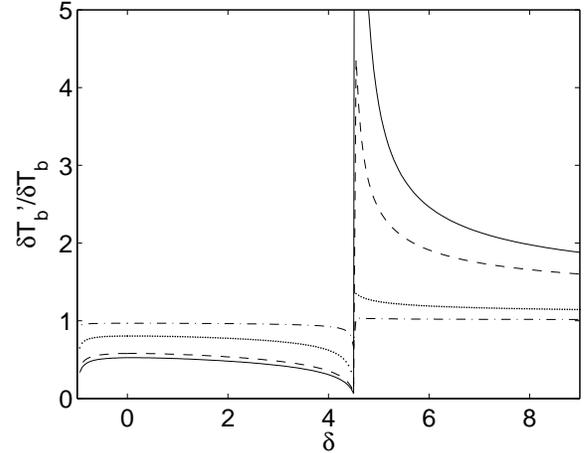}}
\caption{\label{T2}The correction factor to the gas absorption signal for different radiation intensities $y_{\alpha,0}T_{\rm k}=1$ (solid), 10 (dashed), $10^2$ (dotted) and  $10^3$ (dashed-dotted line).}
\end{figure}

A further complication arises from the small-scale density
fluctuations. Even though the resolution of planned future
experiments is currently limited to objects of at least $\sim 1$ Mpc size,
which are still mostly in the linear regime at the relevant epoch, the
spectrum around the resonance is determined by conditions on much
smaller ($<1$ kpc) scale, where nonlinear departures from Hubble flow are important. To illustrate the signal dependence on
the small scale-structure we have plotted the spin temperature  of the adiabatically evolving hydrogen gas at $z=12$ at various overdensities (Figs. \ref{Ga}-\ref{T2}). In general, there is no one-to-one relationship between the density
and the expansion rate, so to make this figure we assumed that all
over/underdensities evolve as spherical ``top-hat'' density perturbations (which explains
the break at $\delta=4.6$ where the gas switches from expansion to
contraction). This yields a unique dependence of $\tau_{\rm GP}$ on local overdensity for a given redshift (Fig. \ref{Ga}). 
The results are shown in Figure \ref{T1} for different values of $y_{\alpha,0}T_{\rm k}$ (i.e. different intensities), which can be related to the ratio of photons in the UV background in the frequency range between Ly$\alpha$ and  Ly-limit per atom, $N_{\alpha}$, according to $ y_{\alpha,0}T_{\rm k}\sim 0.2N_{\alpha}(1+z)^3$.
 The spin temperature changes significantly
when the corrections to the coupling constant (eq. \ref{cor})
are taken into account, except when the radiation
intensity is very high.  
Figure \ref{T2} shows how the local
contributions to the overall 21 cm signal change after including this correction.
The net effect of departures from uniform Hubble expansion on the mean 21 cm signal depends on the weighted contributions from gas at different overdensities, and, since this dependence is non-linear, the contributions of over- and under-dense regions do not cancel each other out.

For hyperfine transition of deuterium, we found that a drastic difference is made by inclusion of Ly$\beta$ photons in the analysis, which practically do not affect hydrogen. Thus the ratio between the deuterium and hydrogen radio signals is not constant, but depends sensitively on the UV spectrum. Unfortunately, this eliminates the possibility of the precision radio measurements of D/H ratio suggested by Sigurdson \& Furlanetto \cite{SF}, except for the period before the first UV sources turn on ($z\gsim 40$).

\begin{acknowledgments}
LC thanks M. Alvarez and A. Loeb for stimulating discussions and McDonald Observatory for the W.J. McDonald Fellowship. We thank S. Furlanetto for helpful comments. This work was supported by NASA 
Astrophysical Theory Program grants NAG5-10825 and
NNG04G177G.
\end{acknowledgments}

\appendix
\section{Energy exchange between photons and atoms}
We consider the scattering of the photon of frequency $\nu$ from the atom approaching with velocity $v$. Let us first ignore the hyperfine splitting. After the scattering the photon energy changes by
\begin{eqnarray}
h\Delta \nu= h\nu (\frac{v}{c}-\frac{h\nu}{mc^2})(1-\cos\alpha_1),
\end{eqnarray}
where  $\alpha_1$ is the angle between direction of the photon before and after the scattering. In the rest frame of the atom, the photon frequency before scattering is $\nu(1-v\cos\alpha_2/c)$, where $\alpha_2$ is  the angle between the direction of the photon and of the atom, so the average loss in energy averaged over all directions and velocities is
\begin{eqnarray}
\label{dis}
\Delta E_1=h\nu \int \phi(\nu(1-v\cos\alpha_2/c))(\frac{v}{c}-\frac{h\nu}{mc^2})
(1-\cos\alpha_1)P(\alpha_1,\alpha_2,v)d\alpha_1 d\alpha_2 dv, \hspace{5mm}
\end{eqnarray}
where $\phi(\nu)$ is the normalized scattering cross-section and $P(\alpha_1,\alpha_2,v)$ is the distribution function. For isotropic radiation field and
Maxwellian velocities integrating \ref{dis} gives
\begin{eqnarray}
\Delta E_1=\frac{(h \nu)^2}{mc^2}(1-\frac{kT_{\rm k}}{h}\frac{\phi'(\nu)}{\phi(\nu)}),
\end{eqnarray}
where we have used $\phi(\nu(1-v/c))\approx \phi(\nu)+\phi'(\nu)\nu v/c$.

Now lets consider the effect of the hyperfine splitting.
Let $b$ be the probability that the electron excited from lower/upper hyperfine level decays after photon reemission to a different hyperfine level. Further let $\sigma_{\rm l}(\nu)$ and   $\sigma_{\rm u}(\nu)$ be the photon absorption cross-section for the atoms at respectively low and upper hyperfine state. Then the average energy loss of a photon in a single scattering caused by change of electrons hyperfine state is
\begin{eqnarray}
\label{hen}
\Delta E_2=bh\nu_{\rm hyp}(\frac{\sigma_{\rm l}e^{h\nu_{\rm hyp}/kT_{\rm s}}-\sigma_{\rm u}}{\sigma_{\rm l}e^{h\nu_{\rm hyp}/kT_{\rm s}}+\sigma_{\rm u}})
\end{eqnarray}
Since the frequency of the hyperfine transition, $\nu_{\rm hyp}$, is very small, we can approximate $\sigma_{\rm u}(\nu)=\sigma_{\rm l}(\nu+\nu_{\rm hyp})\approx\sigma_{\rm l}(\nu)+\nu_{\rm hyp}\sigma_{\rm l}'(\nu)$, $\sigma_{\rm l}'(\nu)/\sigma_{\rm l}(\nu)\approx \phi'(\nu)/\phi(\nu)$ and $e^{h\nu_{\rm hyp}/kT_{\rm s}}\approx 1+h\nu_{\rm hyp}/kT_{\rm s}$. Thus \ref{hen} becomes
\begin{eqnarray}
\Delta E_2= \frac{b(h \nu_{\rm hyp})^2}{2kT_{\rm s}}\left(1-\frac{kT_{\rm s}}{h}\frac{\phi'(\nu)}{\phi(\nu)}\right).
\end{eqnarray}

\section{Resonance profile}
When resonant scattering and Hubble expansion are the only important processes, the spectrum evolves according to
\begin{eqnarray}
\label{res1}
\frac{\partial J(\nu)}{\partial t}=\frac{\partial J(\nu)}{\partial \nu} H\nu +nc\int[\sigma(\nu-\Delta \nu)J(\nu-\Delta \nu)-\sigma(\nu)J(\nu)]P(\Delta \nu)d(\Delta \nu),
\end{eqnarray}
where $P(\Delta \nu)$ is the probability of photons frequency to change by $\Delta \nu$ in a single scattering and $n$ is the scattering atoms density. Neglecting terms of order $(\Delta \nu)^3$
we can rewrite eq. \ref{res1} as
\begin{eqnarray}
\label{res2}
\frac{\partial J(\nu)}{\partial t}=\frac{\partial J(\nu)}{\partial \nu} H\nu +nc[\frac{\partial^2 (J(\nu)\sigma(\nu))}{\partial \nu^2}<\frac{(\Delta \nu)^2}{2}>-\frac{\partial (J(\nu)\sigma(\nu))}{\partial \nu} <\Delta \nu>],
\end{eqnarray}
where $<\Delta \nu>$ and $<(\Delta \nu)^2>$ are the expectation values for  $\Delta \nu$ and $(\Delta \nu)^2$. Based on Appendix A we can show that close to the resonance 
\begin{eqnarray}
<\Delta \nu>&=&\frac{ h\nu_\alpha^2}{mc^2}\left(1-\frac{kT_{\rm k}}{h}\frac{\phi'(\nu)}{\phi(\nu)}\right)+\frac{bh \nu_{\rm hyp}^2}{2kT_{\rm s}}\left(1-\frac{kT_{\rm s}}{h}\frac{\phi'(\nu)}{\phi(\nu)}\right), \\
\label{nuav}
<(\Delta \nu)^2>&=&\left(\frac{2\nu_\alpha^2 kT_{\rm k}}{mc^2}\right)+b\nu_{\rm hyp}^2+\left(\frac{h \nu_\alpha^2 }{mc^2}\right)^2.
\end{eqnarray}
The third term on the right side of the eq. \ref{nuav} can be neglected, since for $T_{\rm k}>1$ K, it is at least $\sim 10^3$ times smaller than the first term.
Since the time photons spend close to the resonance is much shorter than a Hubble time, near the resonance the spectrum evolution can be well approximated by a steady state, i.e., $\partial J(\nu)/\partial t=0$. Thus we can rewrite the eq. \ref {res2} as
\begin{eqnarray}
\label{res3}
\frac{\partial }{\partial \nu}\left(J(\nu)H\nu + \left(\frac{h\nu_\alpha^2}{mc^2}+\frac{bh\nu_{\rm hyp}^2}{2kT_{\rm s}}\right)\sigma(\nu)J(\nu)+\left( \frac{k T_{\rm k}\nu_\alpha^2}{mc^2}+ \frac{b\nu_{\rm hyp}^2}{2}\right)\sigma(\nu)\frac{\partial J(\nu)}{\partial \nu}\right)=0,
\end{eqnarray}
Integrating the above equation and expressing it in dimensionless units gives
\begin{eqnarray}
\label{res4}
\phi(x)J'(x)+2[\eta \phi(x)+\gamma ]J(x)=A,
\end{eqnarray}
where  
 $x=(\nu/\nu_\alpha-1)/(2kT_{\rm k}/mc^2)^{1/2}$, $\eta=(1+w/T_{\rm s})(1+w/T_{\rm k})^{-1}h\nu_\alpha/(2kT_{\rm k}mc^2)^{1/2}$, $w=b\nu_{\rm hyp}^2mc^2/2\nu_\alpha^2 k$, $\gamma=\tau_{\rm GP}^{-1}(1+w/T_{\rm s})^{-1}$ and $\tau_{\rm GP}$ is the Gunn-Peterson optical depth to Ly$\alpha$ resonance scattering. 
When photons are far away from the center of the resonance (i.e., $|x|\gg 1$) their intensity is unaffected by scatterings.
 Thus, since $J'(x)$ and $\phi(x)$ go to zero in the limit $|x|\gg 1$, $A/2\gamma$ must be the UV intensity far away from the center of the resonance.

As expected, if the probability of the spin-flip, $b$, is artificially set to be zero (i.e., the effect of hyperfine transition is neglected), eq. \ref{res4} becomes identical to the one derived by Grachev (1989), based upon Chugai(1987), and later rediscovered by Chen \& Miralda-Escude (2004), based upon Rybicki \& Dell'Antonio (1994).

\end{document}